\documentclass[12pt]{article}
\usepackage[cp1251]{inputenc}
\usepackage{amsmath,amssymb}
\usepackage[dvips]{epsfig}
\usepackage[dvips]{graphics}
\begin{document}
\begin{center}
\Large\textbf{Homogeneous isotropic cosmological models with
pseudoscalar torsion function in Poincare gauge theory of gravity and accelerating Universe}\\
\bigskip
\normalsize A.V. Minkevich$^{1,2}$, A.S. Garkun$^1$ and V.I. Kudin$^3$\\
\medskip
\textit{$^1$Department of Theoretical Physics, Belarussian State University,\\
$^2$Department of Physics and Computer Methods, Warmia and Mazury University in Olsztyn, Poland,\\
$^3$Department of Technical Physics, Belarussian National Technic University}
\end{center}
\begin{center}
\begin{minipage}{0.8\textwidth}
\textbf{Abstract.} The "dark energy" problem is investigated in the framework of the Poincare gauge
theory of gravity in 4-dimensional Riemann-Cartan space-time. By using general expression for
gravitational Lagrangian homogeneous isotropic cosmological models with pseudoscalar torsion
function are built and investigated. It is shown that by certain restrictions on indefinite
parameters of gravitational Lagrangian the cosmological equations at asymptotics contain effective
cosmological constant and can lead to observable acceleration of cosmological expansion. This
effect has geometrical nature and is connected with space-time torsion.
\end{minipage}
\end{center}

\section{Introduction}

Cosmological researches fulfilled during several last decades were fruitful for theoretical and
observational cosmology. The creation of inflationary paradigm, which permits to solve a number of
problems of standard cosmological scenario, in particular, to explain the homogeneity and isotropy
of the Universe at initial stages of cosmological expansion, is important achievement of the early
Universe cosmology [1]. The discovery of the acceleration of cosmological expansion at present
epoch is the most principal achievement of observational cosmology [2]. By using Friedmann
cosmological equations in order to explain accelerating cosmological expansion, the notion of dark
energy (or quintessence) was introduced in cosmology.  According to obtained estimations,
approximately 70\% energy in our Universe is connected with some hypothetical form of gravitating
matter --- ``dark energy'' --- of unknown nature. Recently many investigations devoted to dark
energy problem were carried out (see review [3]). According to widely known opinion, the dark
energy is associated with cosmological term. If the cosmological term is connected with  the vacuum
energy density, it is necessary to explain, why it has the value close to critical energy density
at present epoch.

The present paper is devoted to investigation of the ``dark energy'' problem in the framework of
the Poincare gauge theory of gravity (PGTG), which is natural generalization of Einsteinian general
relativity theory (GR) by applying  the gauge approach to theory of gravitational interaction.
According to PGTG the physical space-time possesses the structure of Riemann-Cartan continuum with
curvature and torsion. By using gravitational Lagrangian of PGTG in general form including both a
scalar curvature and various invariants quadratic in the curvature and torsion tensors (see below),
isotropic cosmology was built and investigated in a number of papers  (see [4--6] and references
herein). As it was shown, the PGTG permits to solve the problem of cosmological singularity of GR:
all solutions of generalized cosmological Friedmann equations (GCFE) for homogeneous isotropic
cosmological models (HICM) deduced in the framework of PGTG are regular in metrics, Hubble
parameter , its time derivative, if gravitating matter satisfies certain physically acceptable
condition at extremely high energy densities and pressures [4]. It is a consequence  of
gravitational repulsion effect, which takes place  at extreme conditions and is connected with
space-time torsion essentially [6]. Properties of discussed HICM in PGTG coincide practically with
that of GR at sufficiently small energy densities, which are much less in comparison with limiting
(maximum) energy density for such models. By including cosmological term of corresponding value to
the GCFE, we can obtain regular cosmological solutions with observable accelerating expansion
stage.  However, like GR, the problem of dark energy in such theory is not solved.

From geometrical point of view, the structure of HICM in PGTG can be more complicated in comparison
with models describing by GCFE. Really, in the case of homogeneous isotropic models the torsion
tensor $S^\lambda{}_{\mu\nu}=-S^\lambda{}_{\nu\mu}$ can have the following non-vanishing components
[7, 8]: $S^1{}_{10}=S^2{}_{20}=S^3{}_{30}=S_{1}(t)$, $\displaystyle S_{123} = S_{231} = S_{312} =
S_{2}(t)\frac{R^3r^2}{\sqrt{1-kr^2}} \sin{\theta}$, where $S_{1}$ and $S_{2}$ are two functions of
time, spatial spherical coordinates are used. The functions $S_{1}$ and $S_{2}$ have different
properties with respect to transformations of spatial inversions, namely, the function $S_{2}(t)$
has pseudoscalar character. The GCFE follow from gravitational equations of PGTG for HICM together
with $S_{2}=0$. Obtained physical consequences of GCFE have principal character. However, it is
necessary to note that gravitational equations of PGTG for HICM have also other solution with
non-vanishing function $S_{2}$. The HICM with two torsion functions are studied below in the frame
of PGTG in connection with the dark energy problem. In Section 2 gravitational equations of PGTG
for HICM with two torsion functions are obtained and investigated. In Section 3 the asymptotics of
cosmological solutions for such models is analyzed.

\section{Homogeneous isotropic models with two torsion functions in PGTG}

At fist let us mention some general relations of the PGTG. Gravitational field is described in the
frame of PGTG by means of the orthonormalized tetrad $h^i{}_\mu$ and anholonomic Lorentz connection
$A^{ik}{}_\mu$ (tetrad and holonomic indices are denoted by latin and greek respectively);
corresponding field strengths are torsion $S^i{}_{\mu\nu}$ and curvature $F^{ik}{}_{\mu\nu}$
tensors defined as
\[
\begin{gathered}
S^i _{\mu \,\nu }  = \partial _{[\nu } \,h^i _{\mu ]}  - h_{k[\mu } A^{ik} _{\nu ]},
\hfill \\
F^{ik} _{\mu \,\nu }  = 2\partial _{[\mu } A^{ik} _{\nu ]}  + 2A^{il} _{[\mu } A^k _{|l\,|\nu ]}.
\hfill
\end{gathered}
\]
We will consider the PGTG based on the following general form of gravitational Lagrangian
\begin{eqnarray}\label{1}
{\cal L}_{\rm G}= h\left[f_0\, F+F^{\alpha\beta\mu\nu}\left(f_1\:F_{\alpha\beta\mu\nu}
                +f_2\: F_{\alpha\mu\beta\nu}+f_3\:F_{\mu\nu\alpha\beta}\right)
        + F^{\mu\nu}\left(f_4\:F_{\mu\nu}+f_5\: F_{\nu\mu}\right)+f_6\:F^2
    \right.
\nonumber \\ \left.
    +S^{\alpha\mu\nu}\left(a_1\:S_{\alpha\mu\nu}+a_2\: S_{\nu\mu\alpha}\right)
    +a_3\:S^\alpha{}_{\mu\alpha}S_\beta{}^{\mu\beta}\right],
\end{eqnarray}
where $h=\det{\left(h^i{}_\mu\right)}$, $F_{\mu\nu}=F^{\alpha}{}_{\mu\alpha\nu}$, $F=F^\mu{}_\mu$,
$f_i$ ($i=1,2,\ldots,6$), $a_k$ ($k=1,2,3$) are indefinite parameters, $f_0=(16\pi G)^{-1}$, $G$ is
Newton's gravitational constant. Gravitational equations of PGTG obtained from the action integral
$ I = \int {\left( {{\cal L}_g + {\cal L}_m } \right)\,} d^4 x$, where  ${\cal L}_m$ is the
Lagrangian of matter, contain the system of 16+24 equations corresponding to gravitational
variables $h^i{}_\mu$ and $A^{ik}{}_\mu$.

Any homogeneous isotropic gravitating system in PGTG is characterized in general case by three
functions of time: the scale factor of Robertson-Walker metrics $R$ and two torsion functions
$S_{1}$ and $S_{2}$. Below by using spherical coordinate system, the tetrad is taken in diagonal
form. Then the curvature tensor has the following non-vanishing tetrad components denoted by means
of the sign \^{} :
\[
\begin{gathered}
  F^{\hat 0\hat 1}{}_{\hat 0\hat 1}  = \,F_{\hat 0\hat 2}{}^{\hat 0\hat 2}
    = F_{\hat 0\hat 3}{}^{\hat 0\hat 3}
    \equiv A_1,
  \qquad
  F_{\hat 1\hat 2}{}^{\hat 1\hat 2}
    = F_{\hat 1\hat 3}{}^{\hat 1\hat 3}  = F_{\hat 2\hat 3}{}^{\hat 2\hat 3}
    \equiv A_2, \hfill \\
  F_{\hat 0\hat 1}{}^{\hat 2\hat 3}  = \,F_{\hat 0\hat 2}{}^{\hat 3\hat 1}
    = F_{\hat 0\hat 3}{}^{\hat 1\hat 2}
    \equiv A_3,
  \qquad
  F_{\hat 3\hat 2}{}^{\hat 0\hat 1} = F_{\hat 1\hat 3}{}^{\hat 0\hat 2}
    = F_{\hat 2\hat 1}{}^{\hat 0\hat 3} \equiv A_4, \hfill \\
\end{gathered}
\]
with
\begin{equation}
\begin{gathered}
A_1=\dot{H}+H^2-2HS_1-2\dot{S}_1,
    \hfill \\
A_{2}  = \frac{k} {{R^2 }} + \left( {H - 2S_1 } \right)^2  - S_2^2,
    \hfill \\
A_{3}  = 2\left( {H - 2S_1 } \right)S_2,
    \hfill \\
A_{4}  = \dot S_2+HS_2,\hfill
\end{gathered}
\end{equation}
where $H=\dot{R}/R $ is the Hubble parameter and a dot denotes the differentiation with respect to
time.

The Bianchi identities in this case are reduced to two following relations:
\begin{equation}
\begin{gathered}
\dot A_{2}  + 2H\left( {A_{2}  - A_{1} } \right) + 4S_1 A_{1}  + 2S_2
    A_{4}  = 0,
\hfill \\
\dot A_{3}  + 2H\left( {A_{3}  - A_{4} } \right) + 4S_1 A_{4}  - 2S_2
    A_{1}  = 0.
\end{gathered}
\end{equation}
By using the gravitational  Lagrangian (1) and the Bianchi identities (3), we obtain the following
system of gravitational equations for HICM
\begin{equation}
a\left( {H - S_1 } \right)S_1  - 2bS_2^2  - 2f_0 A_{2}  + 4f\left( {A_{1}^2 - A_{2}^2 }
\right) + 2q_2 \left( {A_{3}^2  - A_{4}^2 } \right) =  - \frac{\rho } {3},
\end{equation}
\begin{equation}
a\left( {\dot S_1  + 2HS_1  - S_1^2 } \right) - 2bS_2^2  - 2f_0 \left( {2A_{1} + A_{2} }
\right) - 4f\left( {A_{1}^2  - A_{2}^2 } \right) - 2q_2 \left( {A_{3}^2  - A_{4}^2 }
\right) = p,
\end{equation}
\begin{equation}
    f\left[ {\left( {\dot A_{1}  + \dot A_{2} } \right) + 4S_1 \left( {A_{1}  + A_{2} }
        \right)} \right] + q_2\, S_2 A_3+\left(2 f-q_1 \right)S_2A_4 + \left( {f_0  + \frac{a} {8}}
        \right)S_1  = 0,
\end{equation}
\begin{equation}
q_2 \left[ {\left( {\dot A_{3}  + \dot A_{4} } \right) + 4S_1 \left( {A_{3}  + A_{4} }
        \right)} \right] - 4fS_2 A_2-2 \left( q_1+q_2 \right)S_2A_1
    - \left( {f_0  - b} \right)S_2  = 0,
\end{equation} where
\[
\begin{gathered}
  a = 2a_1  + a_2  + 3a_3, \qquad b = a_2  - a_1,
\hfill\\
  f = f_1  + \frac{{f_2 }} {2} + f_3  + f_4  + f_5  + 3f_{6}\, ,
\hfill \\
  q_1  = f_2  - 2f_3  + f_4  + f_5  + 6f_{6}, \qquad q_2  = 2f_1  - f_2 ,
\hfill \\
\end{gathered}
\]
$\rho$ is the energy density, $p$ is the pressure and the average of spin distribution of
gravitating matter is supposed to be equal to zero. Equations (4)--(5) do not contain high
derivatives for the scale factor $R$, if $a=0$, moreover, equations (6)--(7) take more symmetric
form, if $2f=q_1+q_2$. Then the system of gravitational equations for HICM take the following form:
\begin{equation}
 - 2b\,S_2^2  - 2f_0 A_{2}  + 4f\left( {A_{1}^2  - A_{2}^2 } \right) + 2q_2 \left( {A_{3}^2  - A_{4}^2 } \right)
 =  - \frac{1}{3}\rho,
\end{equation}
\begin{equation}
 - 2b\,S_2^2  - 2f_0 \left( {2A_{1}  + A_{2} } \right) - 4f\left( {A_{1}^2  - A_{2}^2 } \right) - 2q_2
    \left( {A_{3}^2  - A_{4}^2 } \right) = p,
\end{equation}
\begin{equation}
    f\left[ {\left( {\dot A_{1}  + \dot A_{2} } \right) + 4S_1 \left( {A_{1}  + A_{2} }
        \right)} \right] + q_2\, S_2 \left( {A_{3}  + A_{4} } \right) + f_0 S_1  = 0,
\end{equation}
\begin{equation}
    q_2 \left[ {\left( {\dot A_{3}  + \dot A_{4} } \right) + 4S_1 \left( {A_{3}  + A_{4} }
        \right)} \right] - 4f\,S_2 \left( {A_{1}  + A_{2} } \right)
    - \left( {f_0  - b} \right)S_2  = 0.
\end{equation}
The system of equations (8)--(11) together with definition of curvature functions (2) is the base
of our investigation of HICM below. Note also that the conservation law for spinless matter has
usual form:
\begin{equation}
    \dot \rho  + 3H\left( {\rho  + p} \right) = 0.
\end{equation} From (8)--(9) follows that
\begin{equation}
    A_{1}  + A_{2}  = \frac{1} {12f_0 }\left( \rho  - 3p\right) - \frac{b}{f_0}S_2^2.
\end{equation}
By using  (13) and the formula following from definition~(2) of curvature functions
\[
    A_{3}^2  - A_{4}^2 = 4A_{2}\, S_2^2  - 4\left( {\frac{k}{{R^2 }} - S_2^2 } \right)S_2^2
        - \left( {\dot S_2  + HS_2 } \right)^2 ,
\]
we find from gravitational equations (8)--(9) the following expressions for $A_1$ and $A_2$:
\begin{eqnarray}
&&
    A_{1}  = -\frac{1} {{12f_0 Z}}\left[ \rho  + 3p - \frac{\alpha } {2}
        \left( {\rho  - 3p - 12bS_2^2 } \right)^2  \right]
        - \frac{\alpha \varepsilon }{Z}\left( {\rho  - 3p - 12bS_2^2 } \right)S_2^2
\nonumber \\
&& \phantom{ A_{1}  = -\frac{1} {{6f_0 Z}} \frac{{\rho  + 3p}} {2}}
        + \frac{{3\alpha \varepsilon f_0 }} {Z}\left[ {\left( {HS_2  + \dot S_2 } \right)^2
        + 4\left( {\frac{k}{{R^2 }} - S_2^2 } \right)S_2^2 } \right],
\\
&&
    A_{2}  = \frac{1} {{6f_0 Z}}\left[ {\rho  - 6bS_2^2  + \frac{\alpha }{4}
            \left( {\rho  - 3p - 12bS_2^2 } \right)^2 } \right]
\nonumber \\
&& \phantom{ A_{1}  = -\frac{1} {{6f_0 Z}} \frac{{\rho  + 3p}} {2}}
        - \frac{{3\alpha \varepsilon f_0 }} {Z}\left[ {\left( {HS_2  + \dot S_2 } \right)^2
        + 4\left( {\frac{k}{{R^2 }} - S_2^2 } \right)S_2^2 } \right],
\end{eqnarray}
where $Z \equiv 1 + \alpha \left[ {\left( {\rho  - 3p} \right) - 12\left( {b + \varepsilon f_0 }
\right)S_2^2 } \right]$, $\alpha  \equiv \frac{f} {{3f_0^2 }}$, $\displaystyle\varepsilon  \equiv \frac{{q_2 }} {f}$,
$\displaystyle\frac{{q_2 }} {{f_0 }} =  3\alpha\, \varepsilon f_0$.  By using the conservation law
(12), the formula (13) and the following relation obtained from
 definition of $A_3$ and $A_4$
\begin{equation}
A_{3}  + A_{4}  = \dot S_2  + 3HS_2  - 4S_1 S_2,
\end{equation}
we obtain from (10) the following expression for the torsion function $S_1$:
\begin{equation}
S_1  = \frac{{3\alpha }} {{4Z}}\left[ 4\left( {2b - \varepsilon f_0 } \right)S_2\,
        \dot S_2 -HY\right],
\end{equation}
where
\[
    Y \equiv \left( {\rho  + p} \right)\left( {3\frac{{dp}} {{d\rho }}} -1 \right) + 12\varepsilon
        f_0 S_2^2.
\]
Then by using formulas (13) and (16) we find from (11) the following differential equation of
second order for the torsion function $S_2$:
\begin{multline}
    \varepsilon \left[
        \ddot S_2  + 3H\dot S_2  +
        \left(3\dot H  - 4\dot S_1\right)S_2
        + 4S_1 S_2\left(3H-4S_1\right)
    \right]
\\ 
        - \left[\frac{1} {{3f_0 }}\left( {\rho  - 3p - 12bS_2^2 } \right)
            + \frac{{\left( {f_0  - b}\right)}} {f}
            \right]S_2  = 0\,.
\end{multline}
The obtained expressions (14)--(15) for curvature functions $A_2$ and $A_1$ together with their
definition (2) give the generalization of cosmological Friedmann equations for HICM:
\begin{multline}
    k/R^2 + (H-2S_1)^2= \frac{1} {{6f_0 Z}}
        \left[
            {\rho  +6\left(f_0 Z- b\right) S_2^2
            + \frac{\alpha }{4} \left( {\rho  - 3p - 12bS_2^2 } \right)^2 }
        \right]
\\
        - \frac{{3\alpha \varepsilon f_0 }} {Z}
            \left[
                {\left( {HS_2  + \dot S_2 } \right)^2
                + 4\left( {\frac{k}{{R^2 }} - S_2^2 } \right)S_2^2 }
            \right],
\end{multline}
\begin{multline}
    \dot{H}+H^2-2HS_1-2\dot{S}_1 = -\frac{1} {{12f_0 Z}}
        \left[
            \rho  + 3p - \frac{\alpha } {2} \left( {\rho  - 3p - 12bS_2^2 } \right)^2
        \right]
        - \frac{\alpha \varepsilon }{Z}\left( {\rho  - 3p - 12bS_2^2 } \right)S_2^2
\\
        + \frac{{3\alpha \varepsilon f_0 }} {Z}
            \left[ {\left( {HS_2  + \dot S_2 } \right)^2
                + 4\left( {\frac{k}{{R^2 }} - S_2^2 } \right)S_2^2 }
            \right].
\end{multline}
 These equations contain the torsion function $S_1$ determined by (17) and the torsion
function $S_2$, satisfying the equation (18). Besides indefinite parameter $\alpha$ determining the
scale of extremely high energy densities [4], obtained equations contain additionally two
indefinite parameters: $b$ with dimension of parameter $f_0$ and the parameter $\epsilon$ without
dimension. We have to analyze all these equations in order to investigate HICM with pseudoscalar
torsion function in the frame of PGTG.

\section{Asymptotics of cosmological solutions for HICM with pseudoscalar torsion function}

The structure of obtained equations describing HICM with pseudoscalar torsion function is
essentially more complicated in comparison with the case of HICM with $S_2=0$. By using the
conservation law (12), the cosmological equation (19) can be written in the following form:
\begin{multline}
 H^2 \left[ {\left( {Z + \frac{{3\alpha }}{2}Y} \right)^2
        + 3\alpha \varepsilon f_0 S_2^2 Z} \right]
    - 6\alpha H\left[ {2\left( {Z + \frac{{3\alpha }} {2}Y} \right)
            \left( {2b - \varepsilon f_0 } \right)S_2 \dot S_2
        - \varepsilon f_0 S_2 \dot S_2 Z} \right]
 \\
    + 36\alpha ^2 \left( {2b - \varepsilon f_0 } \right)^2 S_2^2 \dot S_2^2
    + 3\alpha \varepsilon f_0 \left[ {\dot S_2^2
    + 4\left( {\frac{k} {{R^2 }} - S_2^2 } \right)S_2^2 } \right]Z
    + \left( {\frac{k} {{R^2 }} - S_2^2 } \right)Z^2
 \\
    = \frac{1} {{6f_0 }}\left[ {\rho  - 6b\,S_2^2  + \frac{\alpha } {4}
        \left( {\rho  - 3p - 12b\,S_2^2 } \right)^2 } \right]Z\, .
\end{multline}
From this equation follows, that cosmological solutions possess extreme points for the scale factor
$R$. In fact,  if we put that at $t=0$ the Hubble parameter vanishes $H=0$, then from (21) follows
the relation for $S_2$, its time derivative and energy density, which has to be valid at a bounce:
\begin{multline}
    36\alpha ^2 \left( {2b - \varepsilon f_0 } \right)^2 S_{20}^2 \dot S_{20}^2
    + 3\alpha \varepsilon f_0 \left[ {\dot S_{20}^2
    + 4\left( {\frac{k} {{R_0^2 }} - S_{20}^2 } \right)S_{20}^2 } \right]Z_0
    + \left( {\frac{k} {{R_0^2 }} - S_{20}^2 } \right)Z_0^2
 \\
    = \frac{1} {{6f_0 }}\left[ {\rho_0  - 6b\,S_{20}^2  + \frac{\alpha } {4}
        \left( {\rho_0  - 3p_0 - 12b\,S_{20}^2 } \right)^2 } \right]Z_0\, .
\end{multline}
(Values of various quantities at extreme point are denoted by index $0$). By using the following
equation of state $p=\rho$ and by choosing some initial values for $S_{20}$, $\dot{S}_{20}$,
$\rho_0$ at a bounce,  particular cosmological solution for flat model ($k=0$) presented in Fig.~1
- Fig.~2 was obtained numerically. We see that the pseudoscalar torsion function $S_2$ and the
Hubble parameter $H$ have some non-vanishing values at asymptotics. Below we analyze the following
question: by what restrictions on indefinite parameters $b$ and $\varepsilon$ asymptotical value
for the Hubble parameter corresponds to observable accelerating cosmological expansion?

\begin{figure}[htb!]
\,\hfill
\begin{minipage}{0.48\textwidth}
\centering{
 \includegraphics[width=\linewidth]{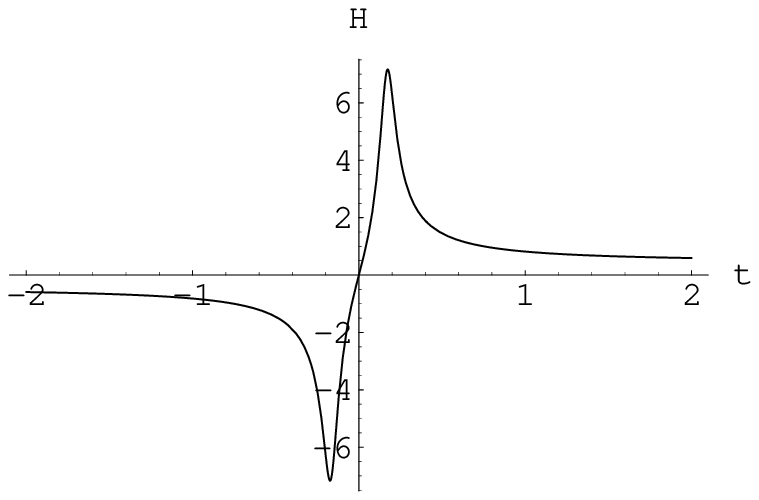}
 \caption{The Hubble parameter $H(t)$ in dimensionless units.}
}
\end{minipage}\, \hfill\,
\begin{minipage}{0.48\textwidth}\centering{
 \includegraphics[width=\linewidth]{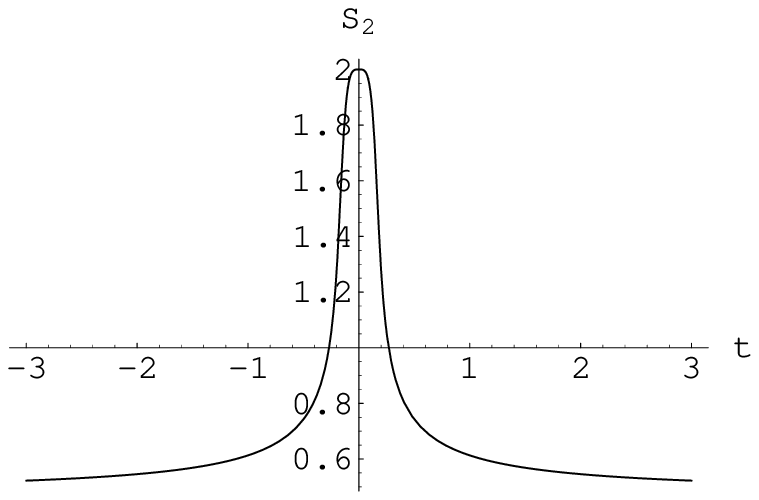}
 \caption{The pseudoscalar torsion function $S_2(t)$ in dimensionless units.}
}
\end{minipage}\,\hfill 
\end{figure}

By taking into account that various parameters of HICM have to be small at asymptotics, we see from
(18), that if $\epsilon\ll 1$, the pseudoscalar torsion function has the following asymptotics:
\begin{equation}
S_2^2  = \frac{{f_0(f_0  - b)}} {{4fb}} + \frac{{\rho  - 3p}} {{12b }}.
\end{equation}
Note, that at asymptotics one uses for gravitating matter the equation of state for dust $p=0$;
then it follows from (23) that $b\leq f_0$ and we obtain at asymptotics: $Z \to (b/f_0)$, $S_1\to
0$. The cosmological equations (19)--(20) at
 asymptotics take the form of cosmological Friedmann equations with cosmological constant:
 \begin{equation}
    \frac{k} {{R^2 }} + H^2  = \frac{1} {{6b }}\left[ {\rho  + \frac{{3\,\left( {f_0  - b} \right)^2}}
         {{4f}}} \right],
\end{equation}
\begin{equation}
    \dot H + H^2  =  - \frac{1} {{12b }}\left[ {\rho  + 3p - \frac{{3\left( {f_0  - b} \right)^2 }}
        {{2f}}} \right].
\end{equation}
 From Eqs.(24)--(25) we see, that parameter $b$ has to be very close to $f_0$.
 The value of $b$ leading to observable acceleration of cosmological expansion  depends on
 the scale of extremely high energy density defined by $\alpha^{-1}$. If we take into account
 that the value of $\frac{3}{4}(f_0-b)^2/f= \frac{1}{4}\alpha^{-1}(1-b/f_0)^2$ has to be equal
 approximately  to $0{.}7\rho_{cr}$  ($\rho_{cr}=6f_0  H_0^2$, $H_0$ is the value of the Hubble
 parameter at present epoch),  then we obtain  that $b=[1-(2{.}8
 \rho_{cr}\alpha)^{1/2}]f_0$.  If we suppose that the value of $\alpha^{-1}$ is greater than the energy
 density for quark-gluon matter, but smaller than the Planckian one, then we can obtain the corresponding
 estimation for $b$, which is extremely close to $f_0$.

\section{Conclusion}

As it was shown, the presence of pseudo-scalar torsion function in HICM built in the framework of
PGTG can lead to effective cosmological constant in asymptotics of cosmological solutions and to
observable accelerating cosmological expansion. The effect of acceleration of cosmological
expansion in PGTG has the geometrical nature and is connected with space-time torsion. From the
point of view of considered theory hypothetical form of gravitating matter --- dark energy or
quintessence --- is fiction.  The further investigation of HICM with pseudoscalar torsion function
will be to continued.

\end{document}